\documentclass{PoS}

\title{Exact beta function and glueball spectrum in large-$N$ Yang-Mills theory}

\ShortTitle{Exact beta function and glueball spectrum in large-$N$ Yang-Mills theory}

\author{\speaker{Marco Bochicchio}\\
        INFN Sezione di Roma \\
Dipartimento di Fisica, Universita' di Roma `La Sapienza' \\
Piazzale Aldo Moro 2 , 00185 Roma  \\
       E-mail: \email{marco.bochicchio@roma1.infn.it}}

\abstract{In the pure large-$N$ Yang-Mills theory there is a quasi-$BPS$ sector that is exactly
 solvable at large $N$. It follows an exact beta function and the glueball spectrum
 in this sector.
 The main technical tool is a new holomorphic loop equation
 for quasi-$BPS$ Wilson loops, that occurs as a non-supersymmetric analogue of Dijkgraaf-Vafa holomorphic
 loop equation for the glueball superpotential of $\cal{N}$ $=1$ $SUSY$ gauge theories.
 The new holomorphic loop equation is localized, i.e. reduced to a critical equation,
 by a deformation of the loop that is a vanishing boundary in homology,
 somehow in analogy with Witten's cohomological localization 
 by a coboundary deformation in $SUSY$ gauge theories.}

\FullConference{European Physical Society Europhysics Conference on High Energy Physics,
EPS-HEP 2009,\\
		 July 16 - 22 2009\\
		 Krakow, Poland}

\begin{document}

% Shorthands for \begin{equation} and the like
 
\def\beq{\begin{equation}}
\def\eeq{\end{equation}}
\def\bea{\begin{eqnarray}}
\def\eea{\end{eqnarray}}
\def\bq{\begin{quote}}
\def\eq{\end{quote}}

\section{Introduction}

The pure $SU(N)$ Yang-Mills ($YM$) theory simplifies
considerably in the large-$N$ limit. For example, to the leading large-$N$ order, the expectation value of a product of normalized
local gauge invariant operators factorizes:
\bea
< \frac{1}{N} \sum_{\alpha \beta} Tr F_{\alpha \beta}^2(x_1)...\frac{1}{N} \sum_{\alpha \beta}Tr F_{\alpha \beta}^2(x_k)>
= 
< \frac{1}{N} \sum_{\alpha \beta} Tr F_{\alpha \beta}^2(x_1)>...
< \frac{1}{N} \sum_{\alpha \beta} Tr F_{\alpha \beta}^2(x_k)>  .
\eea
Thus, to this order, the only information that survives in this correlator is the value of the condensate $< \sum_{\alpha \beta} \frac{1}{N} Tr F_{\alpha \beta}^2(x)>$,
that for a suitable regularization
must be proportional to the appropriate power of the renormalization
group invariant scale, $\Lambda_{QCD}$. In turn $\Lambda_{QCD}$ encodes
the information on the beta function of the large-$N$ theory. 
Another remarkable simplification, that occurs to the next to leading $\frac{1}{N}$
order, is that, because of confinement, the connected two point-functions of local gauge invariant operators
must be saturated by a sum of pure poles. For example, for the scalar glueball
propagator the equation must hold:
\bea
\int < \frac{1}{N} \sum_{\alpha \beta} Tr F_{\alpha \beta}^2(x) \sum_{\alpha \beta} \frac{1}{N} Tr F_{\alpha \beta}^2(0)>_{conn}
\ e^{ip x} d^4x= 
\sum_r \frac{Z_r}{p^2+M_r^2}  .
\eea
The sum of pure poles is constrained by the perturbative operator product 
expansion. It must agree asymptotically for large momentum with the "anomalous dimension"
of the glueball propagator as computed by perturbation theory plus perhaps the sum over the condensates that occur in the operator product expansion  \cite{OP}.
Indeed the scalar glueball propagator behaves in perturbation theory at large momentum, within two-loop accuracy,
up to contact terms, i.e. polynomials in the momentum squared $p^2$, and up to a sum over condensates, as:
\bea
\gamma_G g^4(p) p^4 \log(\frac{p^2}{\mu^2}) ,
\eea
where $\gamma_G$ is a numerical factor \cite{OP}.
The factors of $g$, the renormalized 't Hooft coupling at momentum $p$, occur because of the canonical normalization of the action in perturbation theory and to this order they contain the information
about the one-loop beta function. The extra logarithm in the two-loop computation
is due to an "anomalous dimension". \par
To say it in a nutshell, in this talk we report the computation of a large-$N$ exact beta function \cite{MB} and of 
the glueball propagator in a certain quasi-$BPS$ sector, to be defined in the next sections, of the large-$N$ pure $YM$ theory
to the leading non-trivial $\frac{1}{N}$ order.
We find that the glueball propagator, in the case of Wilsonian normalization of the action,
in the quasi-$BPS$ sector of the large-$N$ theory reduced to two dimensions by the large-$N$ non-commutative Eguchi-Kawai ($EK$) reduction (for a review see \cite{MA} p.6) is:
\bea
\int < \frac{1}{N} Tr(\mu^2)(x_+ x_-)  \frac{1}{N} Tr( \bar \mu^2)(0)>_{conn} e^{i(p_+x_-+ p_-x_+)} d^2x =
\sum_{k=1}^{\infty} \frac{ k^2 \Lambda_W^6}{ \alpha' p_+ p_-+(k \delta-\gamma) \Lambda_W^2} ,
\eea
for some non-vanishing dimensionless coefficients $\alpha',\delta,\gamma$ that admit an expansion in powers of $\frac{1}{\sqrt k}$ and whose precise value will be computed elsewhere.
Here $\mu$, the complex field that occurs in the quasi-$BPS$ sector, is a linear combination of the anti-self-dual
($ASD$) components of the gauge curvature, $\mu=F^-_{01}+iF^-_{02}$. $\bar \mu$ is its Hermitean conjugate and $x_-, x_+$ are light-cone coordinates. $\Lambda_W$
is the renormalization group invariant scale in the Wilsonian scheme.
It is not hard to see that, setting $k^2 \Lambda_W^4 = \frac{1}{\delta^2} [((k \delta-\gamma) \Lambda_W^2+\alpha' p_+ p_-)((k \delta-\gamma) \Lambda_W^2-\alpha' p_+ p_-)+(\alpha' p_+ p_-)^2+(k 2\gamma \delta -\gamma^2) \Lambda_W^4]$, after some simple algebra Eq.(1.4) can be written
as a logarithmic divergent sum that reproduces the correct logarithmic behavior of perturbation theory:
\bea
(\frac{\alpha'}{\delta})^2 \sum_{k=1}^{\infty} \frac{(p_+ p_-)^2 }{ \alpha' p_+ p_- \Lambda_W^{-2}+(k \delta-\gamma)} +... ,
\eea
up to a divergent sum of condensates, proportional to a power of $\Lambda_W$, and up to a divergent sum of contact terms. Eq.(1.4) is rescaled by a power of the 
renormalized canonical coupling constant in the case of canonical normalization of the action, thus accounting of the perturbative dependence on powers
of $g$. It is a result of \cite{MB} that the canonical coupling
renormalizes according to the following 
exact large-$N$ beta function:
\bea
\frac{\partial g}{\partial \log \Lambda}=\frac{-\beta_0 g^3+
\frac {\beta_J}{4} g^3 \frac{\partial \log Z}{\partial  \log \Lambda} }{1- \beta_J g^2 }  ,
\eea
with:
\bea
\beta_0=\frac{1}{(4\pi)^2} \frac{11}{3} , 
\beta_J=\frac{4}{(4\pi)^2} ,
\eea
where $g$ is the 't Hooft canonical coupling constant and $ \frac{\partial \log Z}{\partial \log \Lambda} $ 
is computed to all orders in the 't Hooft Wilsonian coupling constant, $g_W$, by:
\bea
\frac{\partial \log Z}{\partial \log \Lambda} =\frac{ \frac{1}{(4\pi)^2} \frac{10}{3} g_W^2}{1+cg_W^2} ,
\eea
with $c$ a scheme dependent arbitrary constant.
At the same time, the beta function for the 't Hooft Wilsonian coupling
is exactly one loop:
\bea
\frac{\partial g_W}{\partial \log \Lambda}=-\beta_0 g_{W}^3 .
\eea
As a check, once
the result for $ \frac{\partial \log Z}{\partial \log \Lambda} $ to the lowest order in the canonical
coupling, 
\bea
\frac{\partial \log Z}{\partial \log \Lambda}=
\frac{1}{(4\pi)^2} \frac{10}{3} g^2 + ... ,
\eea
is inserted in Eq.(1.6), it implies the correct value of the first and
second perturbative coefficients of the beta function:
\bea
\frac{\partial g}{\partial \log \Lambda}=
-\beta_0 g^3+
(\frac {\beta_J}{4} \frac{1}{(4\pi)^2} \frac{10}{3} -\beta_0 \beta_J) g^5 +...
=-\frac{1}{(4 \pi)^2} \frac{11}{3} g^3 -\frac{1}{(4 \pi)^4} \frac{34}{3} g^5+... ,
\eea
which are known to be universal, i.e. scheme independent. 
It has been argued in \cite{MB} that there is a scheme, that corresponds to a certain choice of $c$, in which the canonical coupling coincides with
a certain definition of the physical effective charge in the inter-quark potential. \par
The plan of this talk is as follows.
The next section on localization summarizes briefly some of the concepts that
are necessary to understand the rationale behind the computation of the beta function and of the glueball spectrum.
This section is very sketchy for reasons of space. See \cite{MB} for more details.
In the last section we describe in the same sketchy way the computation of the glueball spectrum and we outline further
developments, leaving the details for a forthcoming paper.

\section{Localization and holomorphic loop equation}

Extending the finite dimensional theory of Duistermaat-Heckman \cite{DE}, Witten \cite{W1} introduced
localization in the infinite dimensional realm of quantum field theory.
The basic idea of Witten's localization is that the integral of the exponential of 
a closed form, $Q S_{SUSY}=0$, the action, $S_{SUSY}$, of a supersymmetric ($SUSY$) gauge theory in most of the applications,
can be deformed by a coboundary, $Q \alpha$, without changing its value:
\bea
\int \exp(-S_{SUSY}- t Q \alpha) ,
\eea
because $Q^2=0$ and $\int Q\alpha=0$.
Taking the limit $ t \rightarrow +\infty $, the integral localizes on the
set of critical points of the coboundary. 
Thus Witten's localization in quantum field theory is a cohomology
theory in which certain functional integrals are viewed as cohomology classes and they
are computed choosing suitable representatives.
In most of the applications in quantum field theory the cohomology in question
is generated by a $BRST$ operator, $Q$, that in turn is a twisted super-charge of a $SUSY$ gauge theory.
In this case often the $SUSY$ functional integral reduces to the evaluation of a sum of 
finite dimensional integrals  over
the moduli space of instantons. A remarkable example is the partition function of
the $ \cal{N}$ $=2$ $SUSY$ $YM$ theory, whose logarithm is the prepotential. The prepotential
has been found independently on localization arguments by means of the Seiberg-Witten solution.
Later Nekrasov \cite{N} has reproduced the Seiberg-Witten solution using cohomological localization.
A remarkable feature of Nekrasov work is that also the finite-dimensional integrals over the instantons moduli
space can be evaluated
by localization methods provided a compactification of the moduli space of instantons is chosen.
The compactification is absolutely necessary to assure that the integral of a coboundary vanishes,
i.e. to define the integrals over instantons moduli as cohomology classes.
The choice of the compactification introduces to some extent a certain arbitrariness, that is fixed by 
physical or mathematical arguments \cite{N}.
As a result, according to Nekrasov, in the  $\cal{N}$ $=2$ $SUSY$ case, the
functional integral reduces to a sum of Abelian instantons
with no moduli, that can be performed exactly \cite{N}. \par 
Since cohomology is dual to homology, we may wonder as to whether
we can compute functional integrals by homological rather than cohomological deformations.
Were the answer be affirmative, we could get localization without supersymmetry. \par
The natural arena for homological localization in gauge theory, as opposed to cohomological localization,
is the loop equation \cite{MM}. \par
In general the loop equation is the sum of a classical equation of motion and of a quantum term, that involves the contour integral
along the loop.
By homological localization of the loop equation we mean a homological deformation of the loop for which the quantum term
vanishes and
after which the loop equation is reduced to a critical equation for an effective action \cite{MB}.
Hence the needed homological deformation has to satisfy the following two properties. \par 
It has to leave the expectation value of the loop invariant. \par
It has to imply the vanishing of the quantum term in the loop equation, i.e.
of the term that contains the contour integral along the loop. \par
In this homology framework there is a very natural analogue of the operation of adding a coboundary
in cohomology,
that is based on the zig-zag symmetry of Wilson loops.
The zig-zag symmetry is the invariance of a Wilson loop by the addition of a backtracking arc ending with a cusp.
This deformation is a "vanishing boundary" in singular homology. In a regularized version the arc is the boundary of a tiny strip.
While this symmetry holds classically in most of the cases, quantum mechanically
the renormalization process may spoil it, unless a certain fine tuning of the renormalization scheme
occurs. The reason is that in general Wilson loops have perimeter and cuspidal divergences.
The perimeter divergence is linear in the cut-off scale. The cuspidal divergence is logarithmic,
with a coefficient that in turn is divergent for backtracking cusps. To maintain the zig-zag symmetry
the renormalization scheme must be fine tuned in such a way that the cuspidal divergence cancels
that part of the perimeter divergence that is associated to the extra perimeter due
to the addition of backtracking cusps. 
However, in $SUSY$ theories with extended $SUSY$ there are (locally-$BPS$) Wilson loops that have no
perimeter divergence \cite{Gr}. 
Remarkably this property does not follow directly from the extended $SUSY$ but only from
rotational invariance of the parent theory with minimal $SUSY$, whose dimensional reduction
is the daughter theory with extended $SUSY$ \cite{Gr}.
This opens the way to find Wilson loops with similar properties in the pure large-$N$ $YM$ theory.
In this case the quasi-$BPS$ Wilson loops of the daughter theory, the large-$N$ non-commutative $EK$ reduced theory, share the non-renormalization properties with their supersymmetric cousins
because of the four-dimensional rotational invariance \cite{MB}
of the parent large-$N$ $YM$ theory.  
The quasi-$BPS$ Wilson loops are defined as follows \cite{MB}:
\bea
Tr\Psi(C_{ww})=Tr P \exp i \int_{C_{ww}}(A_z+D_u) dz+(A_{\bar z}+ D_{\bar u}) d \bar z ,
\eea
where $D_u=\partial_u+i A_u$ is the covariant derivative of the non-commutative large-$N$ reduced theory along the non-commutative direction $u$. The plane $z, \bar z$ is instead commutative. The loop, $C_{ww}$, starts and ends at the marked point, $w$. The trace in Eq.(2.2)
is over the tensor product of the $U(N)$ Lie algebra and of the infinite dimensional Fock space that defines
the Hilbert space representation of the non-commutative plane $u, \bar u$ \cite{MB, MA, K}. The limit of infinite non-commutativity is understood, being equivalent to the large-$N$ limit of the commutative gauge theory \cite{MA}.
The curvature, $F(B)$, of the connection,
 $B=B_z dz+B_{\bar z} d\bar z=(A_z+D_u) dz+(A_{\bar z}+ D_{\bar u}) d \bar z$, that occurs in the quasi-$BPS$ Wilson loops, is the field $\mu$ that arises in the glueball
propagator in the quasi-$BPS$ sector, $F(B)=\mu=F^-_{01}+iF^-_{02}$.
We stress that the homological localization holds only in the quasi-$BPS$ sector because it is only in this sector
of the large-$N$ $YM$ theory that the zig-zag symmetry is
satisfied without fine tuning of the renormalization scheme.
As a consequence only the glueball propagator for the field $\mu$ can be obtained by homological localization. \par
The second property, the vanishing of the quantum term in the loop equation, is harder to obtain.
In fact this property can be satisfied only if a scheme can be found in which
not only the Wilson loop but also the quantum term in the loop equation
can be regularized in a manifestly zig-zag invariant way.
This is not possible in the usual Makeenko-Migdal loop equation,
not even for quasi-$BPS$ Wilson loops \cite{MB}.
It turns out that it is necessary to write the loop
equation in new variables, changing variables from the connection
to the $ASD$ part of the curvature, introducing in the functional integral the appropriate resolution of identity:
\bea
1= \int \delta(F^{-}_{\alpha \beta}-\mu^{-}_{\alpha \beta}) D\mu^{-}_{\alpha \beta}  \nonumber \\
=\int \delta(F(B)-\mu)|_{\mu=g_-p_+ g_-^{-1}}  D\mu 
\delta(F^{-}_{02}-\mu^-_{02})\delta(F^{-}_{03}-\mu^-_{03}) D\mu^{-}_{02}
D\mu^{-}_{03} ,
\eea
where the integral over $\mu=g p'_+ g^{-1}=g_-p_+ g_-^{-1}$ is on an orbit of the unitary group with measure $D\mu=
\Delta(\mu) Dp_+ Dg_-$, with $g$ unitary, $p'_+$ and $p_+$ upper triangular and $g_-=1+n_-$ with $n_-$ nilpotent and lower triangular.
$\Delta(\mu)$ is the Vandermonde determinant of the eigenvalues of $\mu$.
The resolution of identity that occurs in the right hand side of the second equality contains an integral over the non-Hermitean field, $\mu$, and should be interpreted in the sense of holomorphic matrix models \cite{LA}. This deformation of the Hermitean integral
over $\mu^-_{01}$ to the non-Hermitean complex integral over $\mu$ is absolutely necessary to write the new holomorphic loop equation by means of a further change of variables, defined by the choice of a holomorphic gauge in which $B'_{\bar z}=0$ and $F(B'_z)=\mu'$. This allows us to reduce the quantum term in the loop equation to the evaluation of a residue that can be regularized in a manifestly zig-zag invariant way (see below). \par
In the eighties the change of variables from the gauge connection to the $ASD$ part of the gauge curvature
has been known as the Nicolai map \cite{V}. It has been invented by Nicolai for $SUSY$ gauge theories
because, for example, it leads to remarkable cancellations between the Jacobian of the Nicolai map and
the fermion determinant in the $\cal {N}$ $=1$ $SUSY$ $YM$ theory. However, it makes sense
also in the pure $YM$ theory \cite{MB}, although it does not lead to unexpected cancellations
in this theory. 
The further change of variables to a holomorphic gauge is a new key feature of our approach to the
large-$N$ $YM$ theory. It is based on the idea that quasi-$BPS$ Wilson loops, being holomorphic functionals
of $\mu'$, behave as the chiral (i.e. holomorphic) super-fields of a $\cal{N}$ $=1$ $SUSY$ gauge theory.
In fact the new holomorphic loop equation resembles for the cognoscenti the holomorphic loop equation that occurs
in the Dijkgraaf-Vafa theory of the glueball superpotential in $\cal{N}$ $=1$ $SUSY$ gauge theories \cite{DV,K,LA}.
The partition function thus becomes:
\bea
Z=\int \exp(-\frac{N 8 \pi^2 }{g^2} Q-\frac{N}{4g^2} \sum_{\alpha \neq \beta} \int Tr(\mu^{-2}_{\alpha \beta}) d^4x)
\delta(F^{-}_{02}-\mu^-_{02})\delta(F^{-}_{03}-\mu^-_{03})  
\nonumber \\  \delta(F(B)-\mu)|_{\mu=g_-p_+ g_-^{-1}}   DB D\bar B
\frac{D\mu}{D\mu'}|_{\mu'=G\lambda G^{-1}} D\mu' 
D\mu^{-}_{02}
D\mu^{-}_{03}  ,
\eea
where the integral over $\mu'=G\lambda G^{-1}$ is on an orbit of the complexification of the gauge group with measure
$D\mu'=\Delta(\mu)^2  D\lambda DG$ and with $\lambda$ the diagonal matrix of the eigenvalues of $\mu$.
We can write the partition function in the new form:
\bea
Z=\int \exp(-\frac{N 8 \pi^2 }{g^2} Q-\frac{N}{4g^2} \sum_{\alpha \neq \beta} \int Tr(\mu^{-2}_{\alpha \beta}) d^4x) \nonumber \\
Det'^{-\frac{1}{2}}(-\Delta_A \delta_{\alpha \beta} + D_{\alpha} D_{\beta} +i ad_{\mu^-_{\alpha \beta}} ) 
  \frac{DB^{g_-}}{Dg_- }'    \Delta^{-1}(\mu)   \frac{D(p_+ , g_-)}{D(\lambda,G)} D\mu' D\mu^{-}_{02}
D\mu^{-}_{03} \nonumber \\ = \int \exp(-\Gamma) D\mu' 
D\mu^{-}_{02}D\mu^{-}_{03} ,
\eea
where the integral over the gauge connection of the delta functions has been now explicitly performed
and, by an abuse of notation,
the connection $A$ in the determinants denotes the solution of the equation
$F^-_{\alpha \beta}- \mu ^-_{\alpha \beta}=0$.
The $ ' $ superscript in the first two determinants requires projecting away the zero modes due to gauge invariance
and possibly to moduli, since
gauge-fixing is not yet implied though it is understood.
The determinant of zero modes associated to the (holomorphic) moduli is $ \frac{DB^{g_-}}{Dg_-}'$,
where $B^{g_-}$ is the gauge transform of $B$ by the singular gauge transformation for which $F(B^{g_-} )=p_+$ \footnote{This gauge transformation is in fact non-singular on the Mandelstam graph introduced later to 
obtain localization.}. The new holomorphic loop equation follows
\footnote{  The holomorphic loop equation is written in linear form since it is assumed that the loop $C_{zz}$ is simple, i.e. it has no self-intersections. } :
\bea
<Tr(\frac{\delta \Gamma}{\delta \mu(z)'}\Psi'(C_{zz}))>=
\frac{1}{2 \pi} \int_{C_{zz}} \frac{ dw}{z-w} <Tr\Psi'(C_{zw})> ,
\eea
where $\Psi'$ is the holonomy of $B$ in the gauge $B'_{\bar z}=0$. $\Gamma$ is determined by the holomorphic
loop equation up to an anti-holomorphic functional of $\mu$.
The residue is regularized in a gauge invariant way by analytic continuation from Euclidean to hyperbolic signature
and $\Gamma$ is defined globally on the double cover of the conformal compactification of space-time
with hyperbolic signature \cite{MB,Mas}:
\bea 
\frac{ dw}{z-w}\rightarrow \frac{ dy_+}{x_+-y_+ +i \epsilon} . 
\eea
In a lattice regularization the integral over the $ASD$ curvature would  live
over plaquettes that are dual, in the plane over which the loop lies, to points.
These points become the cusps that are the end points, $p$, of the backtracking strings, $b_p$, that
perform the deformation of the loop, $C$. Adding the backtracking strings implies the homological 
localization of the holomorphic loop equation:
\bea
<Tr(\frac{\delta \Gamma([b_p])}{\delta \mu'(z_p)}\Psi'(C \cup [b_{p}]))>=0 .
\eea
The regularized residue vanishes at the backtracking cusps \cite{MB} because of its manifest zig-zag symmetry.
The lattice regularization of the
Nicolai map is obtained identifying the cusps with parabolic singularities of the reduced $EK$ theory.
These point-like parabolic singularities of the partial large-$N$ $EK$ reduction
are daughters of codimension-two singularities of the four-dimensional
parent gauge theory. The following lattice functional integral is a discretization, corresponding to the lattice of parabolic singularities, of the resolution of identity
that defines the Nicolai map in Eq.(2.3):
\bea
1= \int \delta(F^{-}_{\alpha \beta}(A)-\sum_p \mu^{-}_{\alpha \beta}(p)  \delta^{(2)} (z-z_p(u, \bar u)) ) \prod_p D\mu^{-}_{\alpha \beta}(p) .
\eea
Codimension-two singularities of this kind have been introduced in \cite{MB1} in the pure $YM$ theory as an "elliptic fibration of parabolic bundles" and later in \cite{W2} in the $\cal{N}$ $=4$ $SUSY$ $YM$ theory, for the study of the geometric Langlands correspondence, under the name of "surface operators". In fact they have been studied originally at classical level in \cite{KM} as singular instantons. It turns out that 
when a codimension-two surface is non-commutative, as in our case, the $YM$ action of the corresponding non-commutative
reduced $EK$ model is rescaled by a power of the inverse cut-off  (\cite{MA} p.6 and \cite{K} p.21 ) that cancels precisely \cite{MB} the quadratic
divergence that occurs evaluating the classical $YM$ action on surface operators.
The effective action, $\Gamma([b_p])$, is defined on a Mandelstam graph, that is a conformal transformation of the 
half-plane, obtained drawing backtracking strings ending with pairs of cusps. 
We may think that it is a change of the conformal structure that generates the cusps.
Since the quasi-$BPS$ loops are diagonally embedded in space-time \footnote{This can be seen by the fact that
the connection $B$ contains terms of the form $D_u dz$, implying implicitly that $dz=du$ on the Riemann 
surface over which the loop lies.}, this two-dimensional conformal transformation
lifts to a conformal rescaling of the four-dimensional metric and thus acts by the renormalization group ($RG$) by adding a conformal anomaly to the effective
action, that amounts to a local counterterm, i.e. to change of the subtraction point, $\Gamma([b_p])=\Gamma([p])+ConformalAnomaly([b_p])$. Therefore there is a symmetry of the $RG$ flow 
that generates the homological deformation of the loop by a vanishing boundary, i.e. by backtracking strings.
This is the analogue of the action being a closed form in cohomology, since in the last case there is a symmetry of the action (i.e. the twisted supersymmetry) that generates the coboundary. 
The effective action is a functional defined on a lattice of surface operators. The beta function is obtained extracting the divergences of the effective action. An important point is that a regularization exists for which the loop
expansion of the first functional determinant in Eq.(2.5) satisfies the usual power counting
as in the background-field computation of the beta function. This regularization of the effective action is a point-splitting regularization of the propagator in the background of the lattice of surface operators. A typical example is
the following one-loop logarithmic contribution to the beta function in Euclidean configuration space:
\bea
\frac {1}{ (4 \pi^2)^2 } \sum_{ p \neq p' }  \int d^2u d^2v  \frac{N Tr(\mu_p \bar \mu_{p'}) }{(|z_p-z_{p'}|^2+|u-v|^2)^2} ,
\eea
where the sum over $p, p'$ runs over the planar lattice of the parabolic divisors of the surface operators.
The logarithmic divergence arises for a $p$-independent $\mu_p$.
Had the contribution with $p=p'$ been included, there would appear a quadratic divergence, thus spoiling
the usual power counting. This lattice point-spitting regularization \footnote{This regularization has been found during joint work with Arthur Jaffe.} is followed by Epstein-Glaser renormalization in Euclidean configuration space (see \cite{EG} for references) and it is a possible starting point of a new constructive approach to large-$N$ $YM$ theory.

\section{The glueball spectrum}

It turns out that the beta function is saturated by the $Z_N$ non-Abelian vortices of the
$EK$ reduction \cite{MB}:
$[D_z, D_{\bar z}]-[D_u, D_{\bar u}]= \sum_{p} \mu_p \delta^{(2)}(x-x_p)- H1$,
$[D_{\bar z}, D_{u} ]= 0 $,
$[D_{z}, D_{\bar  u}]=0$. Here $H$ is the (vanishing small) inverse of the parameter of non-commutativity. 
For $Z_N$ vortices of charge $k$, $N-k$ eigenvalues in a $SU(N)$ orbit of the residue of the curvature, $\mu_p$, 
are equal to
$\frac{2 \pi k}{N}$ and $k$ eigenvalues are equal to $\frac{2 \pi (k-N)}{N}$. The complex dimension of the
local moduli space at each point, $p$, which coincides with the complex dimension of the $SU(N)$ orbit, is $k(N-k)$.
From Eq.(2.5) it follows that the effective action in the $\mu$-$\bar \mu$-sector \footnote{This is the effective action whose critical equation
is equivalent to the loop equation for both the connection $B$ and its Hermitean conjugate $\bar B$.
This explains the occurrence of the square of the modulus of the Vandermonde determinant in Eq.(3.1).
Equivalently the effective action in the $\mu$-$\bar \mu$-sector can be obtained by holomorphic/anti-holomorphic fusion \cite{C}, on the double cover of the conformal compactification of space-time with hyperbolic signature \cite{Mas}, inserting the
anti-holomorphic resolution of identity on the image of a hemisphere by the antipodal map.} is given by:
\bea
\Gamma=\frac{N 8 \pi^2 }{g^2} Q+ \int Tr(\frac{N}{g^2}\mu \bar \mu+ \log|\Delta(\mu)|^2 ) d^2x
- \log Det'^{-\frac{1}{2}}(-\Delta_A \delta_{\alpha \beta} + D_{\alpha} D_{\beta} +i ad_{\mu^-_{\alpha \beta}}). 
\eea
The logarithm of the product of the determinant of zero modes and of the determinant for the choice of the holomorphic gauge has been omitted, since it does not contribute to the $\mu$-$\bar \mu$ propagator, because it is the sum of a holomorphic and an anti-holomorphic functional of $\mu$. The logarithm of the Vandermonde determinant is instead
a holomorphic functional everywhere but at coinciding eigenvalues and therefore it must be included in the effective action in the $\mu$-$\bar \mu$-sector.
As an aside we notice that in the $\cal {N} $ $=1$  $SUSY$ $YM$ theory the Jacobian of the Nicolai map would cancel the gluinos determinant in the light-cone gauge \cite{V}, while the Vandermonde determinant would be absent, since
by ordinary cohomological localization due to the tautological Parisi-Sourlas supersymmetry associated to the Nicolai map \footnote{ M. Bochicchio, to appear.}
the partition function (with some insertions of gluinos operators to be non-vanishing)  would be
localized on instantons, for which $\mu^-_{\alpha \beta}=0$, as opposed to vortices. In this case only the topological term (i.e. the second Chern class, $Q$, not to be confused with the $BRST$ charge) and
the logarithm of the (super-)determinant of zero modes due to the instantons moduli, $SDet \omega$, would survive in the effective action.
Of course localization on instantons reproduces the $NSVZ$ beta function \cite{NSVZ}
since the only source of divergences are the instantons zero modes. \par
Since in this talk we do not compute the precise value of the coefficients that occur in the glueball
propagator, but only its general structure, we do not bother about the precise normalization of each
term that occurs in the effective action. 
Working in the Wilsonian scheme the beta function is exactly one-loop.
We want to extract the glueball spectrum from the effective action.
The easiest part to compute is the glueball potential, that originates 
the glueball masses (squared).
After introducing the density of the vortices:
\bea
\rho=N_v^{-1}\sum_p \delta^{(2)}(z-z_p) ,
\eea
the lowest order contribution to the renormalized glueball potential in the loop expansion of the effective action \footnote{The loop expansion is in fact an expansion in powers of the vortices density, $\rho$, that scales as $\frac{1}{\sqrt k}$, see below.} , up to normalization of each term, reads:
\bea
\int \rho^2(x) N Tr(\mu \bar \mu) d^4x+ \int \rho^2(x) \log|\Delta (\mu)|^2 d^4x .
\eea
The subtraction point in each sector labelled by the charge, $k$, of the vortices lattice 
is defined in such a way that the vortices condensate, i.e. the $RG$ invariant part of the gluon
condensate, $\rho^2(x) Tr(\mu \bar \mu)(x)$, is the same in each sector.
This condition is required by general principles, because all the $Z_N$ vortices must "condense at once".
This implies that $\rho^2(x) \frac{k(N-k)}{N}$ is $k$-independent and proportional to $\Lambda_W^4$,
from which remarkable consequences follow. 
Expanding the glueball potential to the second order in $ \mu, \bar \mu $ asymptotically for large $k$, the first term is of order of $\frac{1}{k}$
while the second one is of order of $1$, and both occur with multiplicity $k$.
This occurs because vortices of charge $k$ have eigenvalues with multiplicity $k$ and $N-k$. The contribution of the second term is a distribution in color space of the form $\delta^{(2)}(\mu_i-\mu_j)$, that arises
taking the second derivative of the logarithm of the modulus of the Vandermonde determinant. Its support has non-void intersection with the vortices eigenvalues precisely
because the vortices have an eigenvalues spectrum that is partially degenerated \footnote{The $\delta^{(2)}(0)$
in color space is regularized as prescribed by the large-$N$ $EK$ reduction as $\frac{N}{(2 \pi)^2}$ (see \cite{MA} p.6 and \cite{K} p.21).}.
As we stressed in the previous section, the quantum term in the loop equation is regularized in a gauge
invariant way by analytic continuation from Euclidean to hyperbolic signature \cite{Mas}.
This has remarkable consequences too.
Indeed, after that the renormalization described in the previous section is performed in Euclidean space, the kinetic term for the vortices eigenvalues arises from the finite part of the Jacobian of the Nicolai map
by analytic continuation to hyperbolic signature.
The basic idea is that, after this analytic continuation, the finite parts of the Jacobian can be computed as residues 
in coordinate space by the Cauchy formula. The first term that leads to two derivatives
in the expansion of the Jacobian of the Nicolai map in powers of  $\mu, \bar \mu $ is of order two.
The typical order-two diagram (Eq.(2.10)), after analytic continuation to
hyperbolic signature, leads to \footnote{The restriction to the diagonal $x=u, y=v$ of the fluctuations $\delta \mu$ arises from the diagonal embedding of the quasi-$BPS$ Wilson loops. It can be implemented in Eq.(2.9) adding fluctuations of the form $\sum_p \delta\mu^{-}_{\alpha \beta}(p)  \delta^{(2)} (z-z_p(u, \bar u))$ with $z_p(u, \bar u)=u, \bar z_p(u, \bar u)=\bar u$.
In the stringy version \cite{MB2} which is referred to at the end of this section
the diagonal embedding is substituted by a Lagrangian one.}:
\bea
\frac{N}{k} \int Tr(\delta \mu(x) \delta \bar \mu(y)) \frac{1}{((x_+-y_+)(x_--y_-)+(u_+-v_+)(u_--v_-))^2}|_{x=u,y=v} d^2x d^2y \nonumber \\
\eea
and therefore, applying the Cauchy formula, to a kinetic term of the form:
\bea
\frac{N}{k} \int Tr(\delta \mu (x)\partial_+ \partial _- \delta \bar \mu(x)) d^2x  .
\eea
The final result for the glueball propagator in the quasi-$BPS$ sector
in the Wilsonian scheme, asymptotically for large $k$, is therefore of the form:
\bea
\sum_{k=1}^{\infty} \frac{ k \Lambda_W^6}{ \frac{1}{k}\alpha' p_+ p_-+(\delta-\frac{1}{k}\gamma) \Lambda_W^2} .
\eea
In the canonical scheme the field $\mu$ is rescaled by a power of $g$
and this matches the rescaling in perturbation theory. 
The quasi-$BPS$ glueball masses squared are asymptotically linear in the charge of the vortices lattice, $k$, rather than in the angular momentum, $J$. 
From a qualitative point of view the existence of the magnetic quantum number, $k$,
matches quite well the spectrum found by the numerical lattice computation for $SU(8)$ \cite{NU}.
An interpretation of the numerical data in terms
of Regge trajectories, labeled by $J$, seems instead more complicated \cite{NU}.
Morally this calls for a stringy interpretation in which the string theory dual to pure $YM$ that is the most simple
describes the open strings fluctuations of the magnetic-vortices sheets of the condensate (i.e. of the codimension-two singularities in the language of Langlands duality), rather than the fluctuations of the dual confined electric closed string fluxes. This stringy magnetic description follows by 't Hooft picture of the $YM$ vacuum as a dual superconductor.  
In fact this magnetic string theory actually exists in the quasi-$BPS$ sector, by means of a dual
cohomological twistorial string theory \cite{MB2}, that provides conjecturally the cohomology theory dual to the homological localization. We point out that, conjecturally, the electric string theory, dual to the magnetic one considered in this talk, could be obtained by a solution of the loop equation by means of the geometric Langlands duality 
in the sense of Beilinson-Drinfeld as suggested in \cite{MB3} (see par.(6)).
It is an interesting open problem how to extend the homological approach to localization to large-$N$ $\cal{N}$ $=1$ $SUSY$ $YM$ and to large-$N$ $QCD$. In particular in the $\cal{N}$ $=1$ $SUSY$ $YM$ case the cohomological
and homological approaches should coexist, since they apply to different observables, the gluino condensate
and the $BPS$ Wilson loops respectively.

 \acknowledgments{We thank the organizers of the HEP-EPS 2009 conference, and in particular Antonio Polosa,
for the fruitful atmosphere, the interesting conversations and for inviting us to give this talk.}


\begin{thebibliography}{99}
\bibitem{OP} S. Narison, \emph{Nucl. Phys. Proc. Suppl.} {\bf 64} (1998) 210 [{\tt hep-ph/9710281}].
\bibitem{MB} M. Bochicchio, \emph{JHEP} { \bf 0905} (2009) 116 [{\tt hep-th/0809.4662}].
\bibitem{MA} Yu. M. Makeenko, \emph{The first thirty years of large-$N$ gauge theory}, {\tt hep-th/0407028}.
\bibitem{DE} J. J. Duistermaat, G. J. Heckman, \emph{ Invent. Math.} {\bf 69} (1982) 259.
\bibitem{W1} E. Witten, \emph{J. Geom. Phys.} {\bf 9} (1992) 303 [{\tt hep-th/9204083}].
\bibitem{N} N. A. Nekrasov, proceedings of the \emph{ICM, Beijing}, {\bf 3} (2003) 477 [{\tt hep-th/0306211}].
\bibitem{MM} Yu. M. Makeenko, A. A. Migdal, \emph{ Phys. Lett.} {\bf B 88} (1979) 135;  \emph{ Nucl. Phys.} {\bf B 188} (1981) 269.
\bibitem{Gr} N. Drukker, D. Gross, I. Ooguri, \emph{ Phys. Rev. }{\bf D 60} (1999) 125006
[{\tt hep-th/9904191}].
\bibitem{K} H. Kawai, T. Kuroki, T. Morita, \emph{Dijkgraaf-Vafa theory as large-$N$ reduction}, \emph{Nucl. Phys.}
{\bf B 664} (2003) 185 [{\tt hep-th/0303210}].
\bibitem{LA} C. I. Lazaroiu, \emph{JHEP} {\bf 05} (2003) 044 [{\tt hep-th/0303008}].
\bibitem{V} V. De Alfaro, S. Fubini, G. Furlan, G. Veneziano, \emph{ Phys. Lett.}
{\bf B 142} (1984) 1; \emph{ Nucl. Phys.}
{\bf B 255} (1985) 399.
\bibitem{DV} R. Dijkgraaf, C. Vafa, \emph{A perturbative window into non-perturbative physics}, {\tt hep-th/0208048}.
\bibitem{Mas} L. J. Mason, \emph{Global anti-self-dual Yang-Mills fields in split signature and their
scattering}, {\tt math-phys/0505039}.
\bibitem{MB1} M. Bochicchio, \emph{The large-$N$ limit of $QCD$ and the collective field of the Hitchin fibration},     
 \emph{JHEP} {\bf 9901} (1999) 006 [{\tt hep-th/9810015}].
\bibitem{W2} S. Gukov, E. Witten, \emph{ Gauge theory, ramification and the
geometric Langlands program}, {\tt hep-th/0612073}.
\bibitem{KM} P. B. Kronheimer, T. S. Mrowka, \emph{Topology} {\bf 32} (1993); {\bf 34} (1995).
\bibitem{EG} C. Bergbauer, R. Brunetti, D. Kreimer, \emph{Renormalization and resolution of singularities}, {\tt hep-th/0908.0633}.
\bibitem{C} S. Cecotti, C. Vafa, \emph{Topological antitopological fusion}, \emph{Nucl. Phys.} {\bf B 367} (1991) 359.
\bibitem{NSVZ}  V. Novikov, M. Shifman, A. Vainshtein, V. Zakharov,
\emph{ Phys. Lett.} {\bf B 217} (1989) 103.
\bibitem{NU} H. B. Meyer, \emph{Glueball Regge trajectories}, {\tt hep-lat/0508002}.
\bibitem{MB2} M. Bochicchio, \emph{The Yang-Mills string as the $A$-model on the twistor space of the complex two-dimensional projective space with fluxes and Wilson loops: the beta function}, {\tt hep-th/0811.2547}.
\bibitem{MB3} M. Bochicchio, \emph{JHEP} {\bf 0306} (2003) 026 [{\tt hep-th/0305088}]. 
\end{thebibliography}
\end{document}